\documentclass[aps,twocolumn,showpacs]{revtex4b5}

\usepackage{graphicx} 
\usepackage{dcolumn}
\usepackage{amsmath}

\begin{document}

\title{Spin Liquid Phase in the Pyrochlore Antiferromagnet}

\author{Benjamin Canals}
\email[]{canals@polycnrs-gre.fr}
\homepage[]{http://ln-w3.polycnrs-gre.fr/pageperso/canals/}
\affiliation{Laboratoire Louis N\'eel, CNRS, 25 avenue
des Martyrs, Boite Postale 166, 38042 Grenoble Cedex 9, France}

\author{D. A. Garanin}
\email[]{dgaranin@yahoo.com}
\homepage[]{http://www.mpipks-dresden.mpg.de/~garanin}
\affiliation{Max-Planck-Institut f\"ur Physik komplexer Systeme, 
N\"othnitzer Strasse 38, D-01187 Dresden, Germany}

\date{\today}

\begin{abstract}
\vspace{0.4cm}
Correlation functions (CFs) of the classical
Heisenberg antiferromagnet on the pyrochlore lattice are studied 
by solving exactly the infinite-component spin-vector model.
As in many Fully Frustrated Lattices, the constraint due to the minimization of the
energy and the particular structure based on corner sharing tetrahedra
both contribute to the creation of local degrees of freedom.
The resulting degeneracy destroys any magnetic order at all temperature and we
obtain no sign of criticality, even at $T = 0$.
Calculated neutron scattering 
cross sections have their maxima beyond the first Brillouin
Zone and reproduce experimental results obtained on Y(Sc)Mn$_2$ and
CsCrNiF$_6$ as well as theoretical predictions previously obtained
by classical Monte Carlo simulations.
Evidences for thermal and spatial decoupling of the magnetic modes 
are found so that the magnetic fluctuations in this system can be approximated by 
$S(q,T) \approx f(q) h(T)$.
\end{abstract}

\maketitle

\section{Introduction}

Motivated by the search and understanding of new quantum or classical disordered ground
states in magnetic systems, recent years have seen a renewal of interest in
the properties of frustrated Heisenberg antiferromagnets.
Special attention focused on intrinsically frustrated models, where frustration is
mainly coming from the connectivity of the underlying lattice and not only
from competing interactions.
From this point of view, two lattices distinguished themselves~:~the kagom\'e lattice 
(made of corner sharing triangles) and the pyrochlore lattice (made of corner sharing
tetrahedra).
Experimental and theoretical studies \cite{diep94} related to both lattices
have shown that materials as well as models exhibit unconventionnal magnetic properties, 
involving noncollinear or incommensurate orderings, apparently broken ergodicity 
without structural disorder and in general, deviations from canonical behaviors.
For most geometrically frustrated systems, the absence of ordering at finite temperature
or the tendancy to a reduction of the real interactions seems to be a common characteristic.
An experimental signature of this property is a persistent paramagnetic state down to low
temperature without any sign of ordering \cite{ballelfak96} and sometimes a transition to a frozen state at
low temperature \cite{ginstarajgaugre97} ($T_F << |\Theta_{CW}|$), 
with an unconventional slowing down of the dynamics \cite{uemura941,dunsiger96,uemura942}.
The theoretical understanding of these properties is in progress but it seems that several
properties, expected to be generic, are not systematically reproduced and depend on the
material or the model.
One point is at least widely shared: due to frustration, short range correlations develop
at a temperature much lower than the interactions. This phenomenon has been proven by mean
field analysis \cite{rei91} and classical monte-carlo simulations \cite{rei921,rei922,moecha98}.
On the other hand, there are some properties much more dependent on the studied system.
A major theme of study has been the effect of ``order by disorder'' previously quoted by
Villain and co-workers \cite{villain80} and also by Shender \cite{shender82}.
It was noticed that switching on any source of fluctuation (like thermal or quantum fluctuations) 
has a tendancy to lift the degeneracy between the low lying classical ground states.
Indeed, thermal fluctuations give rise to entropic selection in the kagom\'e case
\cite{chalker92,ritchey93} but this kind of selection was shown to be fluctuation-like
dependent \cite{doucot98}, and sometimes absent \cite{huse92,chandra94}.
Other types of perturbation have been taken into account, such as anisotropy 
\cite{bra94}, structural disorder \cite{shender93}, 
longer range interactions \cite{rei91,harris92}
and quantum fluctuations 
\cite{canals98,sobral97,harris91,zeng91,sachdev92,chubukov92,leche97}, 
leading in general to a magnetic mode selection.

The pyrochlore lattice is a three dimensional arrangement of corner sharing tetrahedra
as depicted in Fig.~\ref{pyr_str}.
According to Lacorre definition of
frustration \cite{lac87}, it is probably the most frustrated structure.
Experimentally, several families of compounds are known to crystallize with this geometry.
The first one is an oxide family with the general formula
A$_2^{3+}$B$_2^{4+}$O$_7$ (Refs. \onlinecite{gardner99,greedan91,dunsiger96}) 
(A is a Rare Earth ion and B a transition metal).
The second one concerns spinels of general formula $AB_2O_4$ or $A_2B_2O_4$
(Refs. \onlinecite{wills99,greedan98,fujiwara98}) (A is a Rare Earth ion and B a transition metal).
The third one concerns fluorides of general formula AB$^{2+}$C$^{3+}$F$_6$ 
(Refs. \onlinecite{harris97,harzintunwanswa94}) (A is an alkali metal, B and C are transition metals).
The last one is the intermetallic Laves Phases compounds of general formula AB$_2$ 
(Refs. \onlinecite{shiga94,shiga88}) (A is a Rare Earth ion and B is a transition metal).
Strictly speaking, the only candidate that can be rigorously described by a localized, uniform
lattice model is the first family, as this system is well ordered with a
uniform distribution of magnetically localized ions.
The second and third sets often possess positional or chemical disorder 
\cite{wills99,harris97} that should be taken into account in the model 
and the last one is better described within an itinerant model \cite{ballac96}.

In this paper, we will focus only on the isotropic classical Heisenberg model on the pyrochlore
lattice, neglecting all other possible contributions that might be necessary for 
an understanding of a peculiar material.
This choice is supported by our hope to deduce intrinsic properties which come 
exclusively from geometrical structure.
Our goal is to compute correlation functions using an approximation
of the Heisenberg model, the infinite component spin-vector model.

\section{Lattice structure and model}

The ``next simplest'' approximation for classical spin systems, which follows the 
Mean Field Approximation (MFA), 
consists in generalizing the Heisenberg Hamiltonian for the $D$-component spin vectors:
%\marginpar{dham}
%
\begin{equation}\label{dham}
{\cal H} = - \frac{1}{2} \sum_{i,j} J_{ij} \, {\bf s}_{i} \cdot {\bf s}_{j}  ,  \qquad |{\bf s}_i|=1
\end{equation}
and taking the limit $D\to\infty$.
In this limit, it is possible to take into account exactly the effect of the
Goldstone modes or would be Goldstone modes, which are the most
important modes in frustrated magnets. 
We will only give the outline of the calculation here and give in a further article
the details of the technique.
Nevertheless, readers who are interested in the full derivation can find a
description of the method in Ref. \onlinecite{garcan99}, where 
the antiferromagnetic model on the kagome lattice was considered.
\begin{figure}[t]
\unitlength1cm
\vspace{-1.1cm}
%\centerline{\includegraphics[width=2.7in]{pyr_str.eps}}
\centerline{\includegraphics[width=8cm]{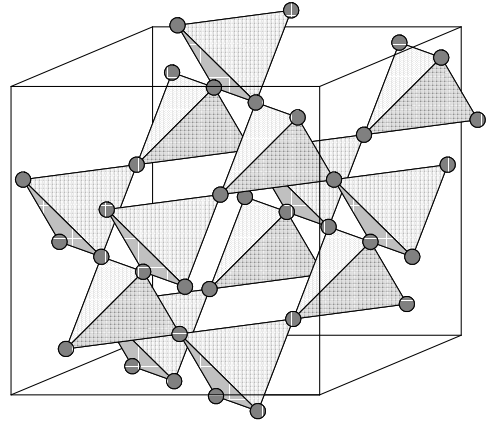}}
\vspace{-4cm}
\caption{ \label{pyr_str} 
Structure of the pyrochlore lattice.
The elementary tetrahedra are labeled by $i=1,\ldots,N$, 
and the sites on tetrahedra (the
sublattices) are labeled by $l=1,2,3,4$.}
\end{figure}
The pyrochlore lattice shown in Fig. \ref{pyr_str} consists of corner-sharing
tetrahedra.
Each node of the underlying Bravais lattice (face centered cubic) is numbered by $i=1,\ldots, N$ and
corresponds to a tetrahedron.
Each site of elementary tetrahedra is labeled by the index $l=1,2,3,4$.
%

%%%%%%%%%%%%%
%%%%%%%%%%%%%

\begin{figure}[t]
\unitlength1cm
\vspace{0cm}
%\centerline{\includegraphics[width=2.7in]{pyr_bri.ps}}
\centerline{\includegraphics[width=7cm]{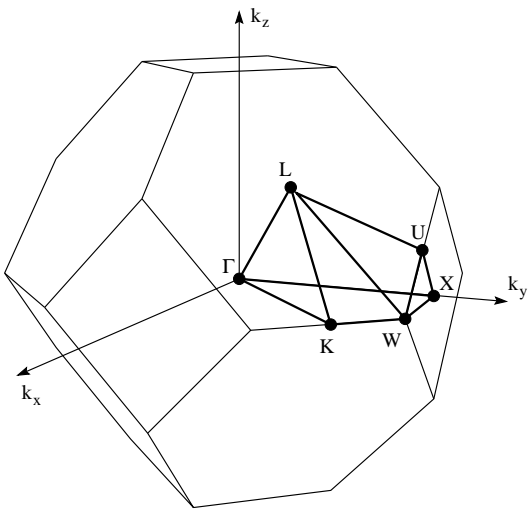}}
\vspace{0cm}
\caption{ \label{pyr_bri} 
Brillouin zone of the pyrochlore lattice.
Letters $\Gamma$, $X$, $W$, $U$ and $L$ are high symmetry points
using the usual crystallographic conventions.}
\end{figure}

It is convenient to put the Hamiltonian (\ref{dham}) into a diagonal form.
First, one goes to the Fourier representation according to
%\marginpar{fourier}
%
\begin{equation}\label{fourier}
{\bf s}_{\bf q}^l = \sum_i {\bf s}_i^l e^{-i {\bf q} r_{i}^{l}}, 
\qquad {\bf s}_i^l = \frac 1N \sum_{\bf q} {\bf s}_{\bf q}^l e^{i {\bf q} r_{i}^{l}}, 
\end{equation}
where the wave vector ${\bf q}$ belongs to the $1^{\rm st}$ Brillouin zone of the
underlying face centered cubic lattice (see Fig. \ref{pyr_bri}).
The Fourier-transformed Hamiltonian reads
%\marginpar{dhamq}
%
\begin{equation}\label{dhamq}
{\cal H} = \frac{1}{2N} \sum_{ll'\bf q} V_{\bf q}^{ll'} {\bf s}_{\bf q}^l 
 \cdot \rm s_{-{\bf q}}^{l'} , 
\end{equation}
where the interaction matrix is given by
%\marginpar{Vq}
%
\begin{eqnarray}\label{Vq}
\hat V_{\bf q} = 2J \nonumber 
\left( \begin{array}{cccc} \nonumber 
0 & b & c & a \nonumber \\
b & 0 & f & d \nonumber \\
c & f & 0 & e \nonumber \\
a & d & e & 0 \nonumber \\
\end{array} \right) , \nonumber
\qquad
\textrm{where} \\
\\
a \equiv \cos( \frac{q_y + q_z}{4})
\qquad
d \equiv \cos( \frac{q_x - q_y}{4}) \nonumber \\
b \equiv \cos( \frac{q_x + q_z}{4})
\qquad
e \equiv \cos( \frac{q_x - q_z}{4}) \nonumber \\
c \equiv \cos( \frac{q_x + q_y}{4})
\qquad
f \equiv \cos( \frac{q_y - q_z}{4}) \nonumber
\end{eqnarray}
At the second stage, the Hamiltonian (\ref{dhamq}) is finally diagonalized to the form
%\marginpar{dhamdiag}
%
\begin{equation}\label{dhamdiag}
{\cal H} =  -\frac{1}{2N} \sum_{n\bf q} \tilde V_{\bf q}^n 
\sigma_{\bf q}^n \cdot \sigma_{-\bf q}^n , 
\end{equation}
where $ \tilde V_{\bf q}^n = 2J \nu_n({\bf q})$ are the eigenvalues of the 
matrix $V_{\bf q}^{ll'} $ taken with a negative sign,
%\marginpar{nudef}
%
\begin{equation}\label{nudef}
\nu_{1,2} = 1, \qquad \nu_{3,4} = - 1 \pm \sqrt{1+Q}.
\end{equation}
where
\begin{eqnarray}\label{Q}
Q=\frac{1}{2} ( & \cos( & \frac{q_y + q_z}{2}) + 
                      \cos( \frac{q_x + q_z}{2}) + \nonumber \\
                & \cos( & \frac{q_x + q_y}{2}) +
                      \cos( \frac{q_x - q_y}{2}) + \\
                & \cos( & \frac{q_x - q_z}{2}) +
                      \cos( \frac{q_y - q_z}{2})      
              ) \nonumber
\end{eqnarray}
The diagonalizing transformation has the explicit form 
%\marginpar{Vtrans}
%
\begin{equation}\label{Vtrans}
U_{nl}^{-1}({\bf q}) V^{ll'}_{\bf q} U_{l'n'}({\bf q}) = \tilde V^n_{\bf q} \delta_{nn'} ,
\end{equation}
where the summation over the repeated indices is implied and $\hat U$ is the real unitary matrix,
$\hat U^{-1} = \hat U^T$, i.e., $U_{nl}^{-1} = U_{ln}$.

The Fourier components of the spins ${\bf s}_{\bf q}^l$ and the 
collective spin variables $\sigma_{\bf q}^n$ are related by
%\marginpar{s-sigma}
%
\begin{equation}\label{s-sigma}
{\bf s}_{\bf q}^l = U_{l n}({\bf q}) \sigma_{\bf q}^n, 
\qquad \sigma_{\bf q}^n = {\bf s}_{\bf q}^l U_{l n}({\bf q}).
\end{equation}
\begin{figure}[t]
\unitlength1cm
\vspace{-0.5cm}
%\centerline{\includegraphics[width=2.7in]{pyr_str.eps}}
\centerline{\includegraphics[width=9cm]{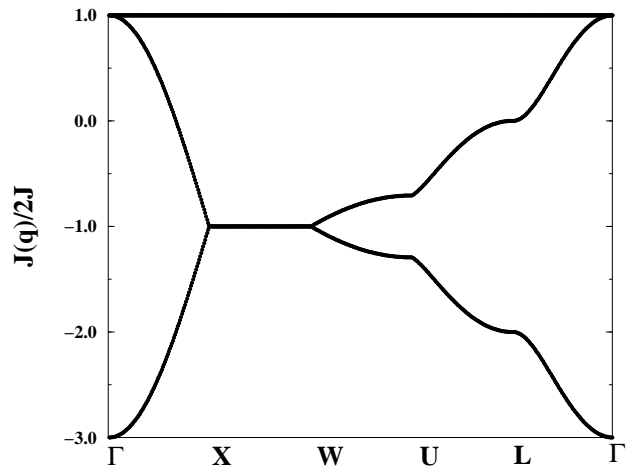}}
\vspace{-0.5cm}
\caption{\label{pyr_mod} 
Reduced eigenvalues of the interaction matrix, $ \nu_n({\bf q}) = \tilde
V_{\bf q}^n/(2J)$ of Eq. (\protect\ref{dhamdiag}). They are plotted such that
they cover the high symmetry axis of the face centered cubic Brillouin
Zone (see Fig. \ref{pyr_bri}).
}
\end{figure}

The largest dispersionless eigenvalues $\nu_{1,2}$ of the interaction matrix 
[see Eq. (\ref{nudef}) and Fig. \ref{pyr_mod}] are a manifestation of frustration in the system 
which precludes an extended short-range order even in the limit $T\to 0$.
Since $\nu_{1,2}$ are independent of ${\bf q}$, 1/2 of all spin degrees of freedom 
are local and can rotate freely.
The eigenvalue $\nu_3$ which becomes degenerate with $\nu_{1,2}$ in the limit
$q\to 0$ is related, to the usual would be Goldstone mode destroying the
long-range order in low-dimensional magnets with a continuous symmetry whereas
the eigenvalue $\nu_4$ is positioned,  in the long-wavelength region, 
much lower than the first two ones, and it is tempting to call it the ``optical'' eigenvalue.

\section{Results}

\begin{figure}
\unitlength1cm
\hspace{0cm}
\centerline{\includegraphics[width=8cm]{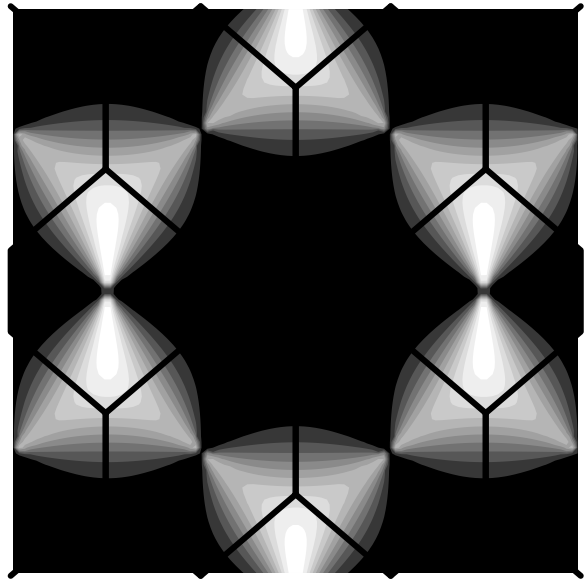}}
\caption{ \label{pyr_neuhex} 
Neutron scattering cross section from the large-$D$ pyrochlore antiferromagnet
at $T=0$ in the ([hh0],[00l]) plane.
(cf. Fig. \protect\ref{pyr_bri}).
Dark regions are low intensity ones while white regions correspond to the
maximum of the scattering.
}
\end{figure}

\begin{figure}
\unitlength1cm
\hspace{0cm}
\centerline{\includegraphics[width=8cm]{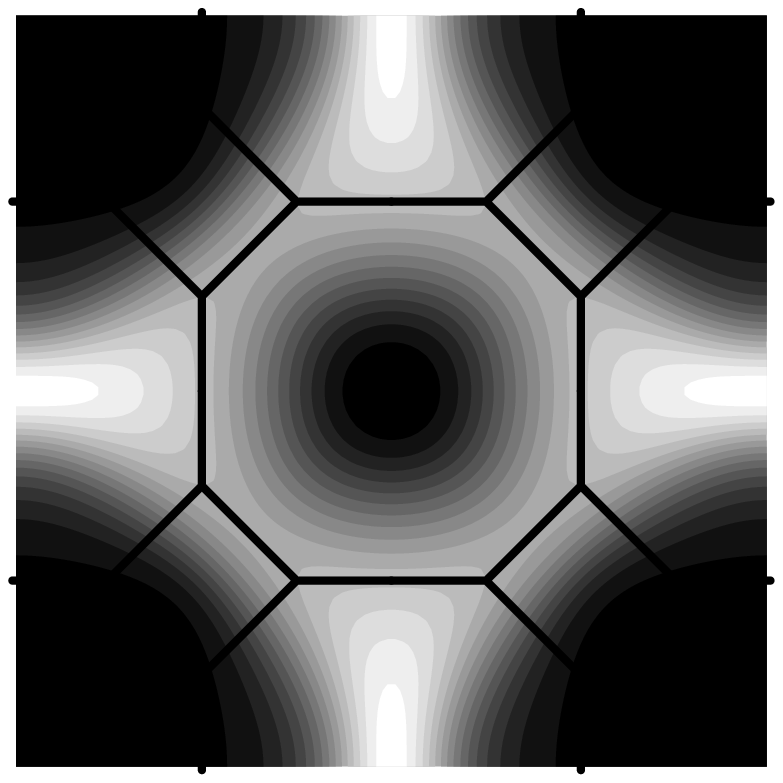}}
\caption{ \label{pyr_neucar} 
Neutron scattering cross section from the large-$D$ pyrochlore antiferromagnet
at $T=0$ in the ([h00],[0l0]) plane.
(cf. Fig. \protect\ref{pyr_bri}).
Dark regions are low intensity ones while white regions correspond to the
maximum of the scattering.
}
\end{figure}

\begin{figure}[t]
\unitlength1cm
\centerline{\includegraphics[width=8cm]{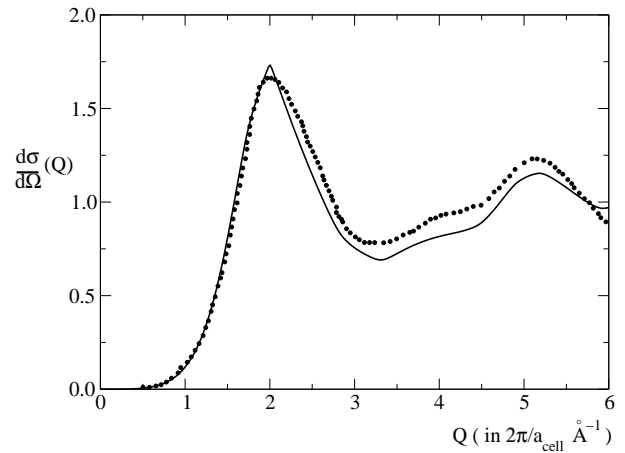}}
\caption{ \label{pyr_powd_comp} 
Powder average of the neutron scattering cross section from the pyrochlore
antiferromagnet. 
The straight line correspond to the $T=0$ case in the $D = \infty$ component classical
antiferromagnet
approach while dots are extracted points from the Monte Carlo results obtained
in Ref.\protect\onlinecite{rei92}
}
\end{figure}

\begin{figure}[t]
\unitlength1cm
\centerline{\includegraphics[width=8cm]{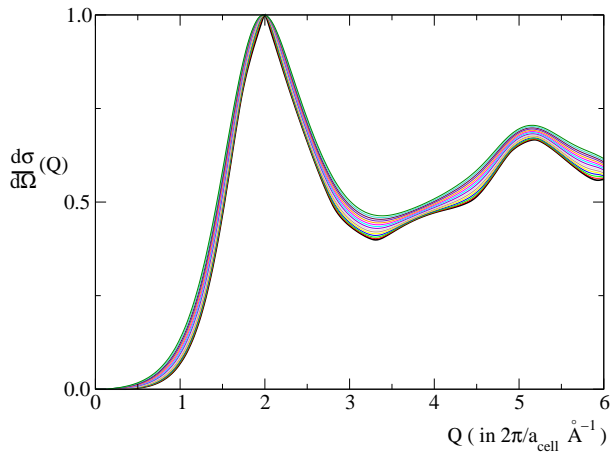}}
\caption{ \label{pyr_powd_multi} 
Powder average of the neutron scattering cross section from the pyrochlore
antiferromagnet at different temperatures ($\theta \in [0,1]$ ).
Each curve is rescaled to be compared with the $T=0$ one.
}
\end{figure}

Following the derivation introduced in Ref. \onlinecite{garcan99}, we compute
all CF's at any temperature
and in particular, the static magnetic neutron scattering cross section :
%\marginpar{defcrsec}
%
\begin{equation}\label{defcrsec}
\frac{ d\sigma }{ d\Omega } \propto \sum_{{\bf rr}'} 
\langle S_{\bf r}^\perp S_{{\bf r}'}^\perp \rangle e^{i{\bf Q}({\bf r} - {\bf r}')},  
\end{equation}
where {\bf Q} is a vector of the reciprocal space and  $r$ and $r^{'}$ design 
spin belonging to the pyrochlore lattice, i.e.,
$S_{\bf r} = S_{i}^{l}$ is a spin located on site $i$ of the Bravais lattice
in the position $l$ of the unit cell.
Two peculiar planes of the reciprocal space have been investigated as shown
in Fig.~\ref{pyr_neuhex} and Fig.~\ref{pyr_neucar}.
The major feature is that there is almost no magnetic signal in the 
first Brillouin zone, all the weight
being recovered in soft cusps outside. In both planes, we observe patterns
very similar to the one obtained by Monte Carlo simulations in 
Ref.~\onlinecite{zinharzei97}
or by neutron diffraction on $Y(Sc)Mn_2$ \cite{ballelfak96}. 
We also note that the cross section in the ([hh0],[00l]) plane is very
close to the one previously calculated in the kagome case by the same
method \cite{garcan99}.
The main implications of these results are that the quantum or classical
nature of the spins in the pyrochlore lattice does not seem to be of
relevance for understanding the liquid nature in terms of
spin-spin correlations.
Indeed, different methods and measures, applied to differents spin
species (classical or quantum), give the same magnetic responses.
Besides, the very strong resemblance between the neutron diffraction
pattern in the ([hh0],[00l]) plane and the one obtained in the kagome
antiferromagnet is evidence that these two lattices are very similar.
This is not surprising as the kagome lattice is a cut of the pyrochlore
lattice perpendicular to the (111) axis.
As previously noted in Ref. \onlinecite{canals98} there are two magnetic modes coexisting 
in this system (see Fig. \ref{pyr_neuhex} and Fig. \ref{pyr_neucar}).
The one related to the ([hh0],[00l]) plane
($Q_1 = [220] \pm [\frac{3}{4} \frac{3}{4} 0]$ or
 $Q_{1}^{'} = [002] \pm [\frac{3}{4} \frac{3}{4} 0]$) would coincide
in an ordered phase with a structure where consecutive tetrahedra are
in phase, while the one in the ([h00],[0l0]) plane 
($Q_2 = [210]$) correspond to a $\pi$ phase between tetrahedra.
Whereas in the quantum approach\cite{canals98} there is a difference of
amplitude between these two modes, within our approach both have
the same amplitude so that the system is unable to select a 
particular one.
%
%This means that the system should be unstable and that a phase transition
%associated to a selection of a mode could be first order.
%
%
The powder average of the neutron cross section 
$\langle d\sigma / d\Omega \rangle$, i.e., the average of 
\ref{defcrsec} over the directions of $q$ is shown 
in Fig. \ref{pyr_powd_comp}.
The continuous line is obtained from the $D = \infty$ component classical
antiferromagnet approach while the dots
correspond to Monte Carlo simulations performed in 
Ref.\onlinecite{rei92}.
There is a good agreement between these two methods and they do not
show any sign of magnetic transition.
We have calculated the effect of the temperature on the powder average
cross section.
In Fig. \ref{pyr_powd_multi} we have plotted the variation of the profiles with reduced
temperature ($\theta = T/T_{\rm MFA}$ where $T_{\rm MFA} = 2J/3$).
One can see that at temperature below $T_{C}^{\rm MFA}$, i.e., below 
$\theta = 1$, these profiles have approximatively the same shape.
In order to control that quantity, we have rescaled each profile and
plotted them one onto the others.
As they almost coincide, it means that at low temperature we can
decouple the $q$ dependance and the $\theta$ dependance so that 
$\langle d\sigma / d\Omega \rangle (Q,\theta) \approx 
 \langle d\sigma / d\Omega \rangle (Q,\theta = 0) f(\theta)$.
The same type of decoupling was assumed in Ref. \onlinecite{canals98} where the authors
proposed that the spatial and the time dependance could be
independent at low temperatures ($\chi (q ,\omega) \approx f(q) g(\omega)$).
These properties are original and certainly come from the high frustration
of the pyrochlore lattice and its related dynamics.
At first approximation, we can state that spatial and time as well as
spatial and thermal
dependance are decoupled at energy scales lower than the mean field
temperature.

\section{Discussion}
\label{secDiscussion}

In this article, we have presented the structure of the neutron diffraction
cross sections obtained from the exact
solution for the $D=\infty$ component classical antiferromagnet on the
pyrochlore lattice.

In contrast to conventionnal three dimensional magnets, there is no ordering
and even no extended short range order at $T=0$.
This anomalous behavior is due to the existence of a would be Goldstone mode
at low temperature but also to the presence of two
dispersionless excitations modes that inhibit any magnetic transition.

The magnetic cross sections are very close to what has been obtained in 
theoretical approaches (Monte Carlo simulations with Heisenberg spins \cite{zinharzei97},
density matrix expansion for spin $S=1/2$ \cite{canals98}) and experimental results
(Y(Sc)Mn$_2$ \cite{ballelfak96}, CsNiCrF$_6$ \cite{harris97,harzintunwanswa94}).
We have reproduced the existence of two dominant magnetic modes that could
explain a first order transition induced by magnetoelastic effects \cite{ballelfak96}.
It is remarkable that the nature of the spins (classical or quantum) as well
as the type of magnetism (localized or itinerant) does not seem to play
a crucial role in the pyrochlore system.
Even the geometry, three dimensional here, seem to have a very little effect on the
struture of the magnetic response.
In fact, we obtain results similar to the ones previously observed on the kagome
lattice.
Even if this point is not solved, it is clear that the close relation between these
two lattices, and the short range scale of the geometry are the relevant 
ingredients in the pyrochlore and the kagome antiferromagnets.

Taking into account assumptions of Ref .\onlinecite{canals98} we can say that
spatial and thermal as well as spatial and time dependences 
of the magnetic cross sections
are decoupled at low energy scale (i.e. $T \lesssim J$) as if 
$S(q,T) \approx f(q) h(T)$ and $S(q,\omega) \approx f(q) g(\omega)$. This striking
result confirms that this model describes an unconventionnal antiferromagnet
and has some similarities with a collective paramagnet.

All these points strongly support that isotropic antiferromagnetic
Heisenberg models (quantum or classical) on the pyrochlore lattice are 
spin liquids.

Although this point is now clearer, it is still an open problem
to understand the experimental results that suggest the appearance of
a topological spin glass like behavior.
There must certainly be other interactions or processes involved in this
case as up to now, no sign of glassiness has ever been seen within theoretical
approaches concerning the Heisenberg model on the pyrochlore lattice.

%\vspace{15mm}

\end{document}